# Social engineering: Concepts, Techniques and Security Countermeasures

Adib Mohammed Syed

*Abstract*— The purpose of this report is to research the topic called Social Engineering in Cyber Security and present the explanation of the meaning, concepts, techniques, and security countermeasures of Social Engineering based on factual academic research.

## I. INTRODUCTION

We live in a technological era where the use of the internet is prevalent this is because the internet is beneficial and vital to all of us in the world. The internet makes our lives easy because internet allows us to communicate using different social media platforms such as WhatsApp, Facebook, and other social media platform. We can also access online banking, shopping and more which means that it saves our time by avoiding travelling to bank or shop [1]. The information or data we input is crucial, the information/data can be private conversation, bank details, and personal medical information. However, there is main issue that the cybercriminal uses the internet to take advantage of innocent people by stealing confidential sensitive information. Cybercriminals use many different techniques for example social engineering and phishing, that are used to carry out the attack on business organisation, people and more [2,3].

Therefore, we need cyber security to deal with those attack techniques as it is the process of protecting data, computer systems including hardware and software, networks and servers from malicious attack, unauthorised access, damage, and the risk of losing the vital confidential data. Without cyber security, our data in computer systems will be at risk of losing to the hands of cybercriminals.

In this report, I will be explaining the concepts of social engineering, how it works and why attackers use this technique to take advantages on people and business organisation. I will also explain the examples of this technique in social engineering and discuss how to prevent social engineers from attacking people and the organisation.

## II. CONCEPTS OF SOCIAL ENGINEERING

The definition of Social Engineering in cyber security is one of the malicious activities for cybercriminals that uses the form of psychological manipulation to manipulate innocent people, which results in them losing their sensitive personal information or data to the attacker [4]. The attacker tends to persuade and gain trust from the victim instead of using any threats, the social engineer depends on the victim's error and this puts the victim in the bad situation. The cybercriminal uses social engineer tactics rather than hacking into the software because it is really easy to manipulate the victim to give the confidential information to the attacker rather than hacking the victim's password to get information in the difficult way [5,6]. The sensitive information can be bank account numbers, passport information, credit and debit card numbers and other data that is confidential and crucial.

The attackers know exactly how to use social engineering effectively and they always have their easy target to aim on victims such as:

Elderly people – The attacker uses communication technology such as email or telephone to communicate with elderly people and deceive them in order to gain access to their confidential data such as bank details. Elderly people are common targets for the attacker as many elderly people are struggling to adapt to using modern technology such as smartphones, laptops, and personal computers. Other reason that could be is that older people often to live at home alone with no people around to protect and support.

Children – Children under the age of 18 years old are also common targets for the attackers as they have lack of knowledge regarding the internet. Most children in the modern era often create online accounts without asking for parent's permission, this means that they are using weak login that the attacker can break in the account easily and accessing to their parent's personal information such as bank details or login details [7]. Another issue that could happen is that the attacker on online such as Facebook or online games easily tricks children, which means the attacker can access to the children's personal information.

Organisation – People and organisations with outdated and weak security such as Firewall, antivirus and other security measured are easy for the attacker to access, the data such as employee's personal data or financial details [8,9]. The reasons could be that the organisation cannot afford to upgrade the security or people in organisation with lack of knowledge can easily give an easy opportunity for the attacker to access by using the techniques of social engineer. Therefore, this will affect the organisation, as they will receive punishment by Data Protection Act 2018 for breaching the law regulation for not protecting the data properly, it can cause them bankruptcy, especially if they are unable to pay the fine as the price of these fines are very high and this can affect organisation's reputation.

Social engineer attacks are effective for targeting exploiting a human's behaviour traits such as excitement, greed, curiosity, fear, anger and sad to give the attacker an opportunity to attack [10]. For example, the attacker sends a fake email to the target stating that "Congrats, you have won £250,000 and to claim the prize, please fill the form with your details.". This will make the target feel excited or greedy, believing that this is legitimate and fills the form that asks for bank details including name, account number and security therefore the attacker will get the bank information from the target. Another example is that a scammer poses itself as the banker sends fake message to the customers that look like they have come from the bank [11]. The fake message to inform the target that their bank account is disabled and asks the target to click the link to restore the bank account. This forces the target to click the link as they are scared and input their ban details. The scammer's goal is getting the target to reply with financial information.

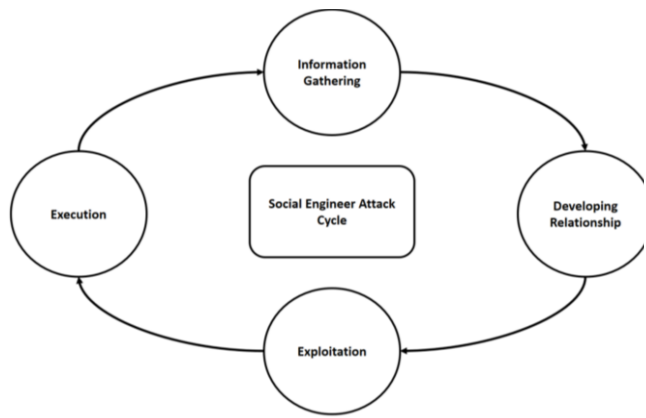

Fig. 1.    Social Engineer Cycle Attack Diagram

As you can see the diagram above that I have created, this is the diagram of social engineer cycle and it is known as the attack cycle [12]. This attack cycle diagram helps the attacker to follow those steps to avoid any confusion and try to achieve their target such as stealing the information or damage the system. It has four stages in social engineer cycle that includes the following, which I will explain each:

Information Gathering – The first step is information gathering and this stage is when the attacker must take their time to research and collect the information about their target such as organisation or people before planning their attack. It could take many weeks or months depending on the situation. The attacker will find the information about their target by researching resources such as on organisation's website or person's online profile from social media like Facebook or Instagram to know more about them. This is the most important stage as this stage allows the attacker to take their time and collect the amount of information for their planning depends on the size of their target such as organisation or people and their security systems so they can assess and identify what routes they could access to their target's data or information [13,14]. The attacker must have accurate information by accessing the information from the organisation or person otherwise, they will not able to succeed if their information is not accurate or incorrect. Therefore, this stage helps to make the attacker becomes comfortable and know how to deal with the target.

Developing relationship – After the attacker collects the correct relevant information about their target, which is helpful and required for their plan to attack, the attacker will have to start developing a relationship with the target. The attacker must be smart and patient with having the conversation with the target to get the target's trust. Otherwise, they will not able to achieve the target's trust. This can be done by the attacker by misrepresenting their fake identification, using relevant collected information, and showing a support for assistance [15]. Therefore, the attacker will do everything to gain the target by creating a conversation with the target on a personal level so this will make the target have more confidence and start believing in the attacker. The communication can be done through emails, social medias, or phone calls [16]. The attacker must take time and maintain the relationship with the target until the trust in between in both parties is settled. For example, this can happen when the attacker builds an online relationship with the target through a fake profile on a social media by sharing pictures, messages, and stories. Once the target trusts the attacker then that is when the attacker will move onto the next step.

Exploitation – The third step is exploitation after the target starts trusting the attacker, this is when the attacker steps up to takes an advantage to use the collected information and relationship to engage with their target to continue to make their relationship stronger [17,18]. The tactical idea of this stage is for the attacker to try encouraging or manipulating the target to reveal the confidential information or data such as login details or bank details without raising any suspicious moves, this term is also called pre-texting. For example, the target reveals bank security number to the attacker on the phone call. After the target give away their sensitive information to the attacker, the security is exposed and left open for the attacker to sneak in easily and gain access to the authorised system.

Execution - Once the attacker has managed to access the system, the attacker will cause a lot of problems for the target such as organisation or individual by attacking them. The attacks can be done by sending them the phishing emails which contains malware harmful files or link or other attachments. At this stage, the attacker should able to succeed its goals such as stealing desired information such as financial details or other confidential details with the help from their target. This will affect the organisation as they lost their sensitive information or data therefore, they will be punished by Data Protection Act 2018 which it will lead them to bankruptcy [19]. Once the attack is accomplished, the attacker must end the interaction with the target and leave the attack scene to avoid leaving any kind of unnecessary suspicious to the target therefore the target is unaware of the attacks. The attacker will remove all the traces of the attack having taken place and will also make sure that they did not leave any evidence or proof behind otherwise their identity and their action will be exposed by the investigators. After the attacker successfully removing any evidence, this will cause the organisations to investigate longer which will lead them to the difficult situation such as fines or reputation damage. This is also giving an advantage to the attacker if their identity is remained hidden and no suspicious occurred then this means that the attacker will return in the future to repeat with using the same process [20].

There are possible reasons why it motivates cybercriminal to use social engineer to attack their target to accomplish their goals such as financial gain, revenge, entertainment, challenge, ego, and espionage.

### III. TECHNIQUES

Social engineer has many techniques that works over email, phone call or social media. The possible reasons why the victims give away their confidential information to the techniques of social engineer easily is due their greed, curiosity, or fear. There are so many techniques that uses in social engineer which I will include the most common used technique and explain the purpose of each technique:

Phishing – This technique is one of most popular social engineer that used in emails or text messages that is aimed to create feeling a sense of fear, panic, greed, can curiosity in victims [21]. When the attacker sends the phishing email or text message to their target which contains malicious link that the takes the target to the fake website such as fake bank website which will ask the target for bank details such as password, account number and security number. This will affect the target as their confidential information will get lost or it sends viruses into the target's device such as smartphone or computer to damage it and steal the information.

Baiting – This technique depends on human's behaviour, such as greed or curiosity which causes the victim to lose their sensitive information [22]. For example, the attacker sends a fake email to the target offering a free iPad and asks them to input their personal information, including bank details. The target loses their money due to baiting. Baiting can be found on emails or text messages or websites. In essence, the attacker offers something the target wants or is interested in and the target falls for it, taking the bait.

Vishing – vishing is one of the techniques in social engineering which does not need internet to manipulate the victims - it is known as voice phishing, which is usually used on phone calls. For example, the attacker poses himself as a banker asking the customer for bank details including security questions and answers, account numbers and security numbers.

Scareware – The purpose of this technique is to make the target feel scared and to manipulate the target to make them believe that their device, such as computer or laptop, is infected by a virus or malware [23,24]. This gives the attacker an opportunity to offer a solution to the target that will fix the fake issues. However, the target will fall for it as the attacker asks the target to download the files that contains malware which gives a breakthrough for the attacker to sneak in and steal information or damage the target's systems. This works for the attacker by using the exchange of pretending to help to fix the target's problems as a way to collect their confidential information.

## IV. SECURITY COUNTERMEASURES

Social engineering attacks can occur to innocent victims as the cybercriminal has these targets to aim at such as elderly people, children, organisation, and other targets over technology, such as email, phone calls or online.

This happened to the sixty-five years old man called Steve who lost two hundred thousand United State dollar (USD) to an investment scam. Steve lived alone at his retirement place; he received a phone call about an investment opportunity. He was not unaware that it was an attacker on the phone, the attacker was acting professional and seemed to have good knowledge of investment, this made Steve believe that it was a legitimate investment. Steve did not have enough funds, so he decided to accept the new investment opportunity [25]. Steve made several transfers to the attacker starting from ten thousand dollars. Steve was given a link to access to the 'professional' website and was asked to create a login account which displayed that his value of money was increasing and the 'market went up' which made Steve to trust the fake website. So, he invested more money up to two hundred thousand dollars. When the 'fake' website went down, he realised that the investment was a scam as he couldn't contact to the attacker on the phone or couldn't access his account. He also did some research and discovered that the fake company was not registered on Australian Securities Investment Commission (ASIC). (Anonymous. Undated. Investment scam: How Steve lost $200 000 to an investment scam. Retrieved from https://www.scamwatch.gov.au/get-help/real-life-stories/investment-scam-how-steve-lost-200-000-to-an-investment-scam)

On this content, I will make a list and explain how to prevent social engineering attacks to avoid losing the sensitive information to the attacker or getting attacked by the attacker.

Do not click on links or attachments in the emails or messages and do not share sensitive information – the email address could be fake but as it is sent from the company such as banks or shops it looks legitimate, but is created by the attacker [26,27]. If you do not know who the sender is, never open the email or never click the link. If you are unsure, the best way is to verify and authenticate with the company. For example, the attacker poses itself as staff member from Amazon sends you a URL link, find the URL link from the official website of Amazon and compare, it will show the difference. Never share your personal information such as password, bank details or phone number and always check the link before clicking on it.

Education, awareness, and training – This method is good for the organisation as it allows to train and educate the employees how to deal with techniques used by social engineers and raise awareness. This will protect the organisation and it can also save the organisation time and money by reducing the risk of the social engineering attack. Set up the security protocols which will explain how to handle data security. Encourage the employees to set up a multi factor authentication that includes password and biometric such as fingerprints, this can be used before payments or before performing a sensitive action. Create an organisational security policy which allows the employees to follow the regulation to avoid any troubles [28].

Secure your devices and service – It is important to keep the device safe. The device can be a laptop, computer, smartphone, etc. Install firewall and antivirus, keep them up to date and scan the system to detect the possible infection. Make sure that software of the device is up to date [29]. Use strong password in logins including longer password, uppercase, numbers and symbols. Install a spam filter that allows to stop every phishing email or messages which reduces the risk of a social engineer attack, the organisation must have a spam filter which will protect their business.

## V. CONCLUSION

Overall, cyber security is more important than ever as the usage of internet is increasing which means cyber attackers using social engineering are also increasing such as phishing, data theft and other attacks. One of the authors mentioned on the book in Chapter 4, page 94 called 'Cyber security culture: counteracting cyber threats through organizational learning and training' which the author states "The journey that we have embarked upon is uncertain and may be more problematic than we first thought; however, our determination and combined strength will ensure that adequate disaster and emergency plans are put in place". I agree with this quote as a user needs to be prepare for the disaster situation and needs to be aware of how to deal with social engineers to protect their personal information.